\documentclass[11pt]{article}
\usepackage[english]{babel}
\usepackage{graphicx}
\usepackage[noadjust]{cite}
\usepackage{amssymb,amsmath,amsthm,mathrsfs}
\usepackage{enumerate}

\newtheorem{theorem}{Theorem}

\theoremstyle{definition}

\theoremstyle{remark}

\newcommand{\tr}{\mathrm{tr}}

\newcommand{\admi}{\mathscr{A}(U)}

\newcommand{\unihh}{\mathcal{U}(\mathcal{H})}
\newcommand{\ope}{A}

\newcommand{\pro}{\boxdot}

\DeclareMathOperator{\jop}{\circ\hspace{-0.3mm}}
\DeclareMathOperator{\lip}{\diamond\hspace{-0.3mm}}
\newcommand{\ranglehs}{\rangle_{\mbox{\tiny HS}}}
\newcommand{\bile}{\hspace{-0.2mm}\big(}
\newcommand{\biri}{\big)}

\newcommand{\spl}{\hspace{1.5mm}}

\newcommand{\hn}{\mathbb{H}_n}
\newcommand{\cent}{\mathcal{Z}}
\newcommand{\centh}{\cent(\hn)}

\newcommand{\modu}{\Delta_G}
\newcommand{\elledue}{\mathrm{L\hspace{-0.2mm}}^{\mbox{\tiny $2$}}}
\newcommand{\elleuno}{\mathrm{L\hspace{-0.2mm}}^{\mbox{\tiny $1$}}}

\newcommand{\hame}{\nu_G}

\newcommand{\ldg}{\elledue(G)}
\newcommand{\elleg}{\elledue(G,\hame;\ccc)}

\newcommand{\lug}{\elleuno(G)}
\newcommand{\elung}{\elleuno(G,\hame;\ccc)}

\newcommand{\ldrn}{\elledue(\mathbb{R}^n)}

\newcommand{\scap}{\langle\cdot\hspace{0.5mm},\cdot\rangle}

\newcommand{\errep}{\mathbb{R}^{\mbox{\tiny $+$}}}

\newcommand{\wt}{\mathcal{W}_\psi}

\newcommand{\cohes}{\left|z\right\rangle}
\newcommand{\pcohes}{\left|z\rangle\langle z\right|}
\newcommand{\fids}{\left|0\right\rangle}
\newcommand{\pfids}{\left|0\rangle\langle 0\right|}
\newcommand{\dispo}{\mathsf{D}}
\newcommand{\qprt}{\frac{1}{\sqrt{2}}(q+\ima\hspace{0.5mm} p)}
\newcommand{\ws}{U}

\newcommand{\unih}{\mathcal{U}(\mathcal{H})}

\newcommand{\reso}{|\ws(g)\psi\rangle\langle\ws(g)\psi|}

\newcommand{\intdisk}{\int_{\mbox{\tiny $|q|^2+|p|^2\le r^2$}}}

\newcommand{\spa}{\hspace{-2mm}}
\newcommand{\mG}{{\nu_G}}

\newcommand{\mm}{\gamma}

\newcommand{\proi}{P}
\newcommand{\fin}{\hspace{0.3mm}}

\newcommand{\tita}{\tilde{\tau}}
\newcommand{\tiq}{\tilde{q}}
\newcommand{\tip}{\tilde{p}}

\newcommand{\screp}{\mathsf{S}}

\newcommand{\erren}{\mathbb{R}^n}
\newcommand{\erredn}{\mathbb{R}^{2n}}

\newcommand{\defi}{\mathrel{\mathop:}=}

\newcommand{\ccc}{\mathbb{C}}

\newcommand{\hh}{\mathcal{H}}

\newcommand{\trc}{\mathcal{B}_{\hspace{-0.1mm}\mbox{\tiny $1$}}(\mathcal{H})}
\newcommand{\hs}{\mathcal{B}_{\mbox{\tiny $2$}}(\mathcal{H})}

\newcommand{\hrho}{\hat{\rho}}
\newcommand{\hsigma}{\hat{\sigma}}
\newcommand{\hpi}{\hat{\pi}}
\newcommand{\htau}{\hat{\upsilon}}

\newcommand{\dequm}{\mathfrak{D}}
\newcommand{\qum}{\mathfrak{Q}}

\newcommand{\kerne}{\mathrm{Ker}\hspace{0.3mm}}
\newcommand{\ran}{\mathrm{Ran}\hspace{0.3mm}}

\newcommand{\norinf}{\|_\infty}

\newcommand{\Hstar}{\mathrm{H}^\ast\hspace{-0.5mm}}
\newcommand{\cinvo}{\hspace{0.2mm}\mathsf{I}\hspace{0.3mm}}
\newcommand{\sinvo}{\hspace{0.2mm}\mathsf{J}\hspace{0.3mm}}

\newcommand{\starp}{\hspace{-0.3mm}\circledast\hspace{-0.3mm}}
\newcommand{\stp}{\hspace{-0.3mm}\star\hspace{-0.3mm}}

\newcommand{\wfsn}{\mathsf{L\hspace{0.2mm}W}_n}
\newcommand{\pwfsn}{\mathsf{W}_n}
\newcommand{\pnwfsn}{\breve{\mathsf{W}}_n}

\newcommand{\cpdn}{\mathsf{P\hspace{-0.5mm}}_n}
\newcommand{\cpdon}{\breve{\mathsf{P}\hspace{-0.5mm}}_n}
\newcommand{\lispn}{\mathsf{LQ}_n}
\newcommand{\qpdn}{\mathsf{Q}_n}
\newcommand{\qpdon}{\breve{\mathsf{Q}}_n}

\newcommand{\qpt}{\hspace{0.2mm}\mathcal{Q}\hspace{0.3mm}}
\newcommand{\obse}{\mathcal{A}}

\newcommand{\de}{\mathrm{d}}
\newcommand{\eee}{\mathrm{e}}

\newcommand{\detz}{\mathrm{d}^{2}\hspace{-0.1mm}z}
\newcommand{\dtn}{\mathrm{d}^{2n}}
\newcommand{\dezn}{\dtn\hspace{-0.2mm}z}
\newcommand{\dezpn}{\dtn\hspace{-0.2mm}z^\prime}

\newcommand{\dqdpn}{\de^n\hspace{-0.3mm}q\hspace{0.9mm}\de^n\hspace{-0.2mm}p}
\newcommand{\dqdpp}{\de^n\hspace{-0.3mm}q^\prime\hspace{0.2mm}\de^n\hspace{-0.2mm}p^\prime}

\newcommand{\intrn}{\int_{\mathbb{R}^n}}
\newcommand{\dxn}{\;\mathrm{d}^n\hspace{-0.2mm}x\hspace{0.4mm}}
\newcommand{\facn}{{\frac{1}{(2\pi)^n}}}

\newcommand{\phasp}{\mathbb{R}^n\hspace{-0.4mm}\times\mathbb{R}^n}
\newcommand{\lrrn}{\elledue(\phasp)}
\newcommand{\lrrnco}{\elledue(\phasp,(2\pi)^{-n}\dqdpn\hspace{0.3mm};\mathbb{C})}

\newcommand{\dug}{\hat{G}}
\newcommand{\cm}{\mathrm{CM\hspace{0.3mm}}}

\newcommand{\pd}{\chi}

\newcommand{\linfg}{\mathrm{L}^{\hspace{-0.4mm}\infty\hspace{-0.3mm}}(G)}

\newcommand{\invo}{^{\ast\hspace{-0.4mm}}}
\newcommand{\ccom}{\mathsf{C}_{\mathsf{c}}}
\newcommand{\gj}{g_j^{-1}}
\newcommand{\gk}{g_k^{\phantom{1}}}

\newcommand{\cast}{\mathrm{C}^\ast}

\newcommand{\tmu}{\widetilde{\mu}}

\newcommand{\zj}{z_j^{\phantom{\ast}}}
\newcommand{\zk}{z_k^{\phantom{\ast}}}

\newcommand{\coj}{\overline{c_j}}
\newcommand{\cok}{c_k}

\newcommand{\tvarrho}{\widetilde{\rho}}

\newcommand{\clat}{\pd_t}
\newcommand{\clas}{\pd_s}
\newcommand{\clats}{\pd_{t+s}}
\newcommand{\claz}{\pd_0}

\newcommand{\csg}{\mathfrak{C}}

\newcommand{\hcsg}{\mathfrak{K}}

\newcommand{\convo}{\hspace{-0.3mm}\circledcirc\hspace{-0.1mm}}

\newcommand{\qp}{{(q,p)}}

\newcommand{\disp}{\exp\!\left(\ima(p\cdot\hq-q\cdot\hp)\right)}
\newcommand{\hq}{{\hat{q}}}
\newcommand{\hp}{{\hat{p}}}
\newcommand{\tq}{\tilde{q}}
\newcommand{\tp}{\tilde{p}}

\newcommand{\fs}{\mathcal{F}_{\hspace{-0.6mm}\mbox{\sf\tiny sp}}}

\newcommand{\intrrn}{\int_{\phasp}\hspace{-0.6mm}}
\newcommand{\intrdn}{\int_{\mathbb{R}^{2n}}}

\newcommand{\ddu}{d_U^{\phantom{1}}}
\newcommand{\duf}{D_U^{\phantom{1}}}
\newcommand{\ee}{\mathrm{e}}

\newcommand{\tre}{\hspace{0.3mm}}
\newcommand{\quattro}{\hspace{0.4mm}}
\newcommand{\cinque}{\hspace{0.5mm}}
\newcommand{\sei}{\hspace{0.6mm}}

\newcommand{\otto}{\hspace{0.8mm}}
\newcommand{\nove}{\hspace{0.9mm}}
\newcommand{\dieci}{\hspace{1mm}}
\newcommand{\dodici}{\hspace{1.2mm}}

\newcommand{\sedici}{\hspace{1.6mm}}
\newcommand{\venti}{\hspace{2mm}}
\newcommand{\mtre}{\hspace{-0.3mm}}

\newcommand{\mcinque}{\hspace{-0.5mm}}
\newcommand{\msei}{\hspace{-0.6mm}}

\newcommand{\erre}{\mathbb{R}}
\newcommand{\toro}{\mathbb{T}}

\newcommand{\cz}{\mathsf{C}_0}

\newcommand{\uni}{\mathcal{U}}

\newcommand{\mut}{\mu_t}

\newcommand{\muno}{\mu_1}
\newcommand{\mdue}{\mu_2}

\newcommand{\ima}{\mathrm{i}}

\newcommand{\intG}{\int_G}

\newcommand{\urep}{{U\hspace{-0.2mm}\vee\hspace{-0.2mm} U}}

\newcommand{\unimu}{{\mut[U]}}

\begin{document}

\title{Discovering the manifold facets of a square integrable representation:
from coherent states to open systems}

\author{
Paolo Aniello$^{1,2}$
\vspace{2mm}
\\ \small \it
$^1$Dipartimento di Fisica ``Ettore Pancini'', Universit\`a di Napoli ``Federico II'',
\\ \small \it
Complesso Universitario di Monte S.~Angelo, via Cintia, I-80126 Napoli, Italy
\vspace{2mm}
\\ \small \it
$^2$Istituto Nazionale di Fisica Nucleare, Sezione di Napoli,
\\ \small \it
Complesso Universitario di Monte S.~Angelo, via Cintia, I-80126 Napoli, Italy
}

\date{}

\maketitle

\begin{abstract}
\noindent  Group representations play a central role in theoretical physics. In particular, in quantum mechanics
unitary --- or, in general, projective unitary --- representations implement the action of an
abstract symmetry group on physical states and observables. More specifically, a major role
is played by the so-called \emph{square integrable representations}. Indeed, the properties of these representations
are fundamental in the definition of certain families of generalized coherent states, in the phase-space
formulation of quantum mechanics and the associated star product formalism, in the definition of an interesting notion of
function of quantum positive type, and in some recent applications to the theory of open quantum systems
and to quantum information.
\end{abstract}


\section{Introduction}


Symmetries and group representations are fundamental in modern science.
E.g., due to Wigner's theorem on symmetry transformations~\cite{WignerGT,Bargmann,AnielloSW,AnielloSW-bis},
(projective) unitary group representations~\cite{Raja,Folland-AA} play a central role in quantum theory.
More specifically, several important topics in theoretical physics and applied mathematics
--- phase-space quantum mechanics, quantization, signal and image processing,
various group-theoretical aspects of quantum information science and of the theory
of open quantum systems etc.\ --- ultimately rely on a remarkable mathematical tool:
the notion of \emph{square integrable representation} of a locally compact
group~\cite{Godement-I,Godement-II,Duflo,Phillips,Grossmann1,AnielloSDP,AnielloPR,Aniello-new}.
In the present contribution, we will discuss --- without any purpose of completeness, but trying to
illustrate how varied the whole subject is --- some interesting examples
where square integrable representations play a fundamental role. The reader may find further examples
in the reference books~\cite{FollandHAPS,DaubechiesTLW,Schroeck,Ali,Fuhr,Gazeau}.

The paper is organized as follows.
In sect.~\ref{sect2}, we first recall that the coherent states of the harmonic oscillator are generated
by the action on a `fiducial vector' of a square integrable projective representation:
the so-called \emph{Weyl system}. We then argue that the usefulness of square integrable representations is mainly
due to certain `orthogonality relations' generalizing \emph{Schur's orthogonality relations} for compact
topological groups~\cite{Folland-AA} . This generalization is highly nontrivial in the case where the relevant group is
\emph{not} unimodular. Moreover, in the case of a square integrable representation which is genuinely projective,
passing to a \emph{central extension} of the relevant group --- so achieving a standard unitary
representation of the central extension --- one obtains, in general, a representation that is square integrable
\emph{modulo the center} only. We also recall that, by means of a square integrable representation, one can define
a linear isometry --- sometimes called the (generalized) \emph{wavelet transform} ---
mapping the carrier Hilbert space of the representation into the Hilbert space of square integrable functions
on the relevant group, and enjoying nice properties.
Standard wavelet analysis involves the (non-unimodular) affine group of the real line.

By means of a square integrable representation --- see sect.~\ref{sect3} --- one can also construct
a pair of \emph{quantization} and \emph{dequantization} maps. Of course, quantizing is traditionally
regarded as an essential step for switching from the classical picture to the quantum setting~\cite{FollandHAPS,Schroeck,Ali,Weyl}.
We stress, however, that the latter map should not be regarded as the `poor sister' of the former:
Dequantizing one obtains a remarkable formulation of quantum mechanics in terms of complex functions, where
the composition of operators is replaced by a `non-local' (i.e., non-pointwise) \emph{star product} of functions.
Specifically, the harmonic analysis associated with a square integrable projective
representation of the group of translations on phase space --- the Weyl system ---
allows one to capture, in a very elegant and effective way, some peculiarities of a quantum system
versus a classical one.

This field of research is still in constant progress and there is room for new investigations that do not fall
within the traditional range of applications of abstract harmonic analysis to theoretical physics; consider,
e.g., some new applications to the theory of open quantum systems. See sect.~\ref{sect4}.

Finally, in sect.~\ref{sect5}, we argue that square integrable representations are an essential ingredient
in the construction of certain \emph{state-preserving products} of trace class operators.


\section{Generalized coherent states and square integrable representations}
\label{sect2}


Recall that the \emph{coherent states} $\{\cohes\}_{z\in\ccc}\subset\elledue(\erre)$ of the quantum harmonic
oscillator~\cite{Ali,Gazeau,AnielloCS} are generated by a family of unitary operators
$\{\dispo (z)\}_{z\in\ccc}$, the so-called \emph{displacement operators}:
\begin{equation} \label{displ}
\cohes = \dispo (z) \fids , \ \ \ z=\qprt\in\ccc \fin ;
\end{equation}
here, the \emph{fiducial vector} $\fids$ is the ground state of the harmonic oscillator Hamiltonian.
The vectors $\{\cohes\}_{z\in\ccc}$ form a tight (continuous) \emph{frame}~\cite{DaubechiesTLW,Ali,Gazeau,AnielloFT};
i.e., they give rise to an integral \emph{resolution of the identity} of the form
\begin{equation}
\frac{1}{\pi}\int \msei\detz \nove \pcohes = I \fin .
\end{equation}

We stress that the displacement operators form a \emph{projective representation}~\cite{Raja}
--- $z\mapsto\dispo (z)$ --- which is often called the \emph{Weyl system}~\cite{AnielloFT,AnielloSP,Aniello-WS}.
Adopting phase-space coordinates $q,p$ --- see~(\ref{displ}) --- and considering the general case of $2n$ degrees of freedom,
the Weyl system is given by
\begin{equation} \label{defws}
G=\erre^n\times\erre^n \ni \qp\mapsto\ws\qp\defi\disp \fin .
\end{equation}
Here, $G=\erre^n\times\erre^n$ is the (additive) group of phase-space translations, and $\hq$, $\hp$ are
the \emph{position} and \emph{momentum} operators in $\elledue(\erren)$ (we will always set $\hbar=1$),
respectively; moreover:
\begin{equation} \label{multipli}
\ws(q + \tq, p + \tp)= \eee^{\frac{\ima}{2}(q\cdot\tp - p\cdot\tq)}\dieci
\ws(q,p)\otto \ws(\tq, \tp) \fin .
\end{equation}
This formula is intimately related to the canonical commutation relations, as it is
clear from the rigorous expression \emph{\`a la} Weyl of these relations~\cite{Folland-AA,Aniello-WS}.
Note that the \emph{multiplier}~\cite{Raja} of $\ws$ --- i.e., the
function $(q,p;\tq,\tp)\mapsto\eee^{\frac{\ima}{2}(q\cdot\tp-p\cdot\tq)}$ on
the rhs of~(\ref{multipli}) --- entails the standard symplectic form in $\erredn$.
Hence, it is \emph{not exact}, so that the representation $\ws$ is \emph{genuinely projective};
namely, it cannot be `converted' into a standard unitary representation of the same group~\cite{Raja,Aniello-WS}.
Accordingly, another important property of the Weyl system --- it is an \emph{irreducible}
representation of an \emph{abelian} group in an \emph{infinite-dimensional} Hilbert space ---
is only compatible with the fact that $U$ is genuinely projective
(all irreducible unitary representations of abelian groups are characters~\cite{Folland-AA}).

On the other hand, by a standard procedure~\cite{Raja}, one can replace the projective representation
$\ws$ with a \emph{unitary} representation $\screp$ of a \emph{non-abelian} group; i.e., the \emph{central extension}
$\hn$ of $G$, the so-called \emph{Heisenberg-Weyl group}~\cite{Folland-AA,FollandHAPS}. This is the group
$\erre\times\erre^n\times\erre^n$, with composition law
\begin{equation}
(\tau,q,p)\sei (\tita,\tiq,\tip) =
(\tau+\tita +(q\cdot \tip - p\cdot \tiq)/2 , q+ \tiq, p+ \tip) \fin ,
\ \ \ \tau,\tita\in\erre, \ q,\tiq, p,\tip\in\erre^n \fin.
\end{equation}
Precisely, one stipulates that $\ws(q,p) = \screp(0,q,p)$, where $\screp$ is
an irreducible unitary representation of the extended group $\hn$, the
\emph{Schr\"odinger representation}~\cite{Folland-AA,FollandHAPS}. It turns out that
\begin{equation} \label{schreps}
\big(\screp (\tau,q,p)\tre f\big)(x)\defi \eee^{-\ima\tre\tau} \big(U (q,p)\tre f\big)(x)
= \eee^{-\ima\tre(\tau+q\cdot p/2)}\sei\eee^{\ima \cinque p\cdot x} f(x-q) \fin ,\ \ \
f\in \elledue(\erren) \fin .
\end{equation}

As already noted, the coherent states generate a resolution of the identity, i.e.,
$U$ is such that
\begin{equation} \label{resuni}
\frac{1}{(2\pi)^n}\int \msei \dqdpn \venti \ws\qp\pfids\ws\qp^\ast = I \fin ;
\end{equation}
here, the fiducial vector $\fids$ may actually be replaced with any other (normalized) nonzero vector.

The integral decomposition~(\ref{resuni}) holds true because the projective representation $\ws$
is \emph{square integrable} or, equivalently, the related unitary representation $\screp$
is \emph{square integrable modulo the center} $\centh=\{(\tau,0,0)\in\hn\colon \tau\in\erre\}$
of $\hn$~\cite{AnielloPR}. Moreover, the possibility of replacing the fiducial vector $\fids$
with \emph{any} other normalized vector is a consequence of the fact that the group $\erre^n\times\erre^n$
is abelian (hence, \emph{unimodular}).

Let us now denote by $U$ a \emph{generic} irreducible --- in general, projective --- representation of
a \emph{locally compact group} $G$, with multiplier $\mm\colon G\times G\rightarrow\toro$,
acting in a separable complex Hilbert space $\mathcal{H}$.
We assume the scalar product $\scap$ in $\hh$ to be linear in its \emph{second} argument.
We denote by $\unihh$ the \emph{unitary group} of $\hh$, by $\hame$ (a normalization of)
the \emph{left Haar measure} on $G$, and by $\modu$ the
\emph{modular function} on $G$~\cite{Raja,Folland-AA}.
For $\psi,\phi\in\mathcal{H}$, consider the bounded and continuous `coefficient' function
\begin{equation} \label{coef}
c_{\psi\phi} \colon G\ni g\mapsto \langle U(g)\,\psi ,
\phi\rangle\in\ccc \fin .
\end{equation}
Functions of this form allow us to define the set
\begin{equation}
\admi\defi\left\{\psi\in\mathcal{H}\,|\
\exists\tre\phi\in\mathcal{H}\colon\ \phi\neq 0,\; c_{\psi\phi} \in \elleg\right\}
\end{equation}
of all \emph{admissible vectors} for $U$. The representation $U$ is called \emph{square integrable} if
$\admi\neq\{0\}$.

Square integrable representations are ruled by the following fundamental
result~\cite{Godement-I,Godement-II,Duflo,Phillips,Grossmann1,AnielloSDP,AnielloPR}:

\begin{theorem} \label{Duflo-Moore}
Let $U\colon G\rightarrow \mathcal{U}(\mathcal{H})$ be a square integrable projective representation.
The set $\admi$ of all admissible vectors is a dense linear subspace of $\hh$,
stable under the action of $U$. For every pair of vectors $\phi\in\mathcal{H}$ and $\psi\in\admi$,
the coefficient $c_{\psi\phi}\colon G\rightarrow\ccc$ is square integrable wrt the Haar measure $\hame$.
Moreover, there exists a unique positive selfadjoint, injective linear operator $\duf$ in $\mathcal{H}$
--- the so-called \emph{Duflo-Moore operator} --- such that $\admi=\mathrm{Dom}\big(\duf\big)$,
and satisfying the orthogonality relations
\begin{eqnarray}
\int_G \overline{c_{\psi_1\phi_1}(g)} \sedici c_{\psi_2\phi_2}(g)\
\de\hame (g)  =  \langle\phi_1 ,\phi_2\rangle\, \langle
\duf\psi_2, \duf\psi_1\rangle \fin ,
\end{eqnarray}
for all $\phi_1,\phi_2\in\mathcal{H}$ and all $\psi_1,\psi_2\in\admi$.
$\duf$ is bounded if and only if $G$ is unimodular --- i.e., $\modu\equiv 1$ --- and, in such case,
it is a multiple of the identity: $\duf = \ddu \sei I$, $\ddu>0$.
\end{theorem}

A few comments are in order:

\begin{enumerate}

\item The square-integrability of a representation extends to its unitary equivalence class.

\item Every irreducible unitary representation of a \emph{compact} group is
square integrable, as the Haar measure of such a group is \emph{finite}.
If this measure is normalized as a probability measure,
then the Duflo-Moore operator is of the form $\dim(\mathcal{H})^{-1/2}\tre I$
(Peter-Weyl theorem~\cite{Folland-AA}).

\item A locally compact group having a non-compact center does not admit
square integrable \emph{unitary} representations~\cite{AnielloPR}.
It may admit square integrable \emph{projective} representations, or unitary representations
that are square integrable modulo the center; see the example of the group
of translations on phase space with the Weyl system, or of the Heisenberg-Weyl group
with the Schr\"odinger representation. The fact that various groups of interest
in physics, like the Poincar\'e group, do not admit square integrable representations
has motivated the investigation of useful alternative approaches~\cite{AnielloPR,Schroeck,Ali,Aniello-impr,AnielloEXT}.

\item  If $U\colon G\rightarrow \mathcal{U}(\mathcal{H})$ is a square integrable
representation, then, for every $\psi\in\mathcal{H}$ such that $0\neq\psi\in\admi$,
the following resolution of the identity holds (compare with~(\ref{resuni})):
\begin{equation}
\|\duf\psi\|^{-2} \int_G \de\hame (g) \dieci \reso = I \fin .
\end{equation}
Moreover, one can define the linear \emph{isometry}
\begin{equation}
\wt\colon \hh\ni\phi\mapsto \|\duf\psi\|^{-2} \tre c_{\psi\phi}\in\elleg \fin ,
\end{equation}
where $c_{\psi\phi}$ is the coefficient function~(\ref{coef}). This map is called
the \emph{generalized wavelet transform} associated with $U$, with \emph{analyzing vector}
$\psi$~\cite{AnielloPR,Ali,Aniello-discr}.

\item The closed subspace $\ran(\wt)\subset\elleg$ is a
\emph{reproducing kernel Hilbert space}~\cite{AnielloPR,Ali,AnielloFT,Aronszajn}, and $\wt$ intertwines
the representation $U$ with the \emph{left regular $\mm$-representation}~\cite{AnielloPR}.
If $U$ is a \emph{unitary} representation, the latter is nothing but
\emph{left regular representation}~\cite{Folland-AA}.

\item The square integrable representations of a \emph{semidirect product}, with an abelian normal factor,
can be classified~\cite{AnielloSDP}. The classical example is the one-dimensional
affine group, which gives rise to the standard wavelet transform~\cite{AnielloSDP,Aniello-new,DaubechiesTLW,AnielloSP}.
In this case, the analyzing vector $\psi$, or \emph{mother wavelet}, must belong to the domain
of the (unbounded) Duflo-Moore operator.

\item In wavelet analysis, one often deals with \emph{discrete} frames,
instead of \emph{continuous} ones~\cite{DaubechiesTLW}.
The possibility of achieving discrete frames from group representations,
and the relation between the existence of such frames and the square-integrability
of the representations, is an important issue, the so-called
\emph{discretization problem}~\cite{Ali,Aniello-discr,Aniello-discr-bis}.

\item Square integrable unitary representations are also known as representations of the \emph{discrete series}
since they appear as \emph{discrete summands} in the integral decomposition
into irreducibles of the left regular representation of a locally compact group~\cite{Folland-AA,Duflo,Fuhr}.

\end{enumerate}


\section{Quantum mechanics `on phase space' and square integrable representations}
\label{sect3}


In addition to the wavelet transform, which maps Hilbert space vectors into complex functions
on a group, by means of a square integrable representation one can also define --- following
various possible approaches --- a pair $(\qum,\dequm)$ formed by a quantization and by a \emph{de}quantization
map~\cite{FollandHAPS,Schroeck,Ali,AnielloFT,AnielloSP}.
These maps transform functions into operators and \emph{vice versa}.
Among the various approaches, we focus on the most natural group-theoretical generalization of the
classical scheme proposed by Weyl, Wigner, Groenewold and Moyal~\cite{Weyl,Wigner,Groenewold,Moyal}.

For the sake of simplicity, here we consider the case
where the relevant group $G$ is unimodular and, in particular,
the special case of the group of translations on phase space.
Let $\hs$ be the Hilbert space of Hilbert-Schmidt operators in $\hh$,
and $\trc\subset\hs$ the Banach space of trace class operators.
Given a square integrable projective representation $U\colon G\rightarrow\uni(\hh)$
(with $G$ \emph{unimodular}), we call \emph{dequantization map} the linear isometry determined by~\cite{AnielloFT,AnielloSP}
\begin{equation}
\dequm\colon\hs\rightarrow\ldg\equiv\elleg \fin , \ \ \
(\dequm\tre\ope)(g) =d_U^{-1}\,\tr(U(g)^\ast\ope) \fin ,
\ \forall\quattro\ope\in\trc \fin ,
\end{equation}
where $\ddu$ is the positive constant appearing in Theorem~\ref{Duflo-Moore},
and we have exploited the fact that $\trc$ is a dense linear subspace of the Hilbert space $\hs$.
The \emph{quantization map} associated with $U$ is nothing but the Hilbert space adjoint
of the dequantization map; namely, the surjective partial isometry
\begin{equation}
\qum \defi \dequm^\ast\colon\ldg\rightarrow\hs \fin .
\end{equation}
Clearly, the linear map $\qum$ has, in general, a nontrivial kernel $\kerne(\qum)=\ran(\dequm)^\perp$.

The \emph{star product}~\cite{AnielloSP,AnielloRSP} associated with the pair $(\qum,\dequm)$ is the binary operation defined by
\begin{equation} \label{defstp}
\ldg\times\ldg\ni (f_1, f_2)\mapsto f_1\stp f_2 \defi \dequm ((\qum f_1) \tre (\qum f_2))\in\ldg \fin .
\end{equation}
For functions living in $\ran(\dequm)$ this operation can be regarded as the \emph{dequantized product}
of operators. The pair $(\ldg,\tre\stp\tre)$ is a $\Hstar$-\emph{algebra}~\cite{AnielloSP,Ambrose},
whose \emph{annihilator ideal} is $\ran(\dequm)^\perp$.

For this group-theoretical star product we have, among others~\cite{AnielloSP,AnielloRSP}, the following result:

\begin{theorem}[\cite{AnielloSP}]
Let $U\colon G\rightarrow\unih$ --- $G$ unimodular --- be a square integrable projective representation,
with multiplier $\mm$: $U(gh)=\mm(g,h)\sei U(g)U(h)$. For every $f_1,f_2\in\ldg$, we have:
\begin{eqnarray}
\big(f_1\stp f_2\big)(g) \spa & = & \spa d_U^{-1} \intG f_1(h) \sei
\big(\proi f_2\big)(h^{-1}g) \dieci \overline{\mm (h,h^{-1}g)} \dodici \de\mG (h)
\nonumber \\ & = & \spa
d_U^{-1} \intG \big(\proi f_1\big)(h)\sei f_2(h^{-1}g) \dodici \overline{\mm (h,h^{-1}g)}
\dieci \de\mG (h)
\nonumber \\
\label{threeeqs} & = & \spa
d_U^{-1} \intG \big(\proi f_1\big)(h)\sei \big(\proi f_2\big)(h^{-1}g)
\dieci \overline{\mm (h,h^{-1}g)} \dodici \de\mG (h) \fin ,
\end{eqnarray}
where $\proi$ is the orthogonal projector in $\ldg$ onto $\ran(\dequm)$.
Thus, for every $f_1,f_2\in\ran(\dequm)$,
\begin{equation} \label{fstpruni}
\big(f_1\stp f_2\big)(g) = d_U^{-1} \intG f_1(h) \sei
f_2(h^{-1}g) \, \overline{\mm (h,h^{-1}g)} \dodici \de\mG (h)
\fin  .  \ \ \ \mbox{\rm (`$\mm$-twisted convolution')}
\end{equation}
\end{theorem}

In the case of the group of translations on phase space --- $G=\phasp$ --- $\hh=\ldrn$,
$U$ is the Weyl system, $\ran(\dequm)=\ldg=\lrrnco$ (i.e., the annihilator ideal is trivial)
and $\ddu=1$. Taking into account the fact that
$\mm(q,p \cinque ; q^\prime,p^\prime) = \exp(\ima(q\cdot p^\prime - p\cdot q^\prime)/2)$,
the $\mm$-twisted convolution is nothing but the classical \emph{twisted convolution}~\cite{FollandHAPS,AnielloSP}.
Note however that the function $(\dequm\tre\hrho)\qp=\tr(U(q,p)^\ast\hrho)$
associated with a \emph{quantum state} --- $\hrho\in\trc$, $\hrho\ge 0$, $\tr(\hrho)=1$ ---
is \emph{not} its \emph{Wigner distribution} $\rho$~\cite{FollandHAPS,Wigner},
but rather the corresponding \emph{quantum characteristic function} $\tvarrho$.
Let us clarify this point by first recalling the notion of characteristic function in classical probability theory
and then discussing its quantum counterpart.

Recall that endowing the Banach space $\lug\equiv\elung$ (where, now, $G$ is a generic locally compact group)
with the \emph{convolution product} and with a suitable \emph{involution} --- i.e., considering the triple
\begin{equation*}
\big(\lug, \convo, \cinvo\colon\varphi\mapsto\varphi\invo\big) \fin , \
(\varphi_1\convo\varphi_2)(g) \defi \intG \varphi_1(h)\tre\varphi_2(h^{-1}g) \; \de\hame(h) \fin ,
\ \varphi\invo(g) \defi \modu(g^{-1})\dieci \overline{\varphi(g^{-1})}
\end{equation*}
--- we get a \emph{Banach $\ast$-algebra}, the `group algebra'~\cite{Folland-AA,AnielloPF,AnielloFPT}.
A positive, bounded linear functional on the $\ast$-algebra $\big(\lug, \convo, \cinvo\big)$
--- i.e., a suitable function in $\linfg$ --- is called a \emph{function of positive type} on $G$~\cite{Folland-AA};
namely, $\pd\in\linfg$ is of positive type if
\begin{equation} \label{clapo}
\intG \pd (g) \otto (\varphi\invo \convo \varphi) (g) \; \de\hame(g)\ge 0 \fin ,
\ \ \ \forall\quattro\varphi \in \lug \fin . \ \ \
\mbox{(PTF condition)}
\end{equation}
A function of positive type $\pd\in\linfg$ agrees $\hame$-almost everywhere with a bounded continuous
function~\cite{Folland-AA}, the `continuous version' of $\pd$, and
\begin{equation} \label{nori}
\|\pd\norinf = \pd(e)
\ \ \ \mbox{($\pd(e)\equiv$ value at the identity $e\in G$ of the `continuous version' of $\pd$)} .
\end{equation}
Moreover, for a \emph{bounded continuous} function $\pd\colon G\rightarrow\ccc$ the following facts are
equivalent~\cite{Folland-AA,AnielloPF}:
\begin{description}

\item[P1\spl] $\pd$ is of positive type;

\item[P2\spl] $\pd$ is such that
\begin{equation} \label{clapo-bis}
\intG \intG \pd (g^{-1}h) \cinque \overline{\varphi(g)}\quattro \varphi(h) \; \de\hame(g)\quattro\de\hame(h)\ge 0 \fin ,
\ \ \ \forall\quattro\varphi \in \ccom(G) \fin ;
\end{equation}

\item[P3\spl] for every finite set $\{g_1,\ldots,g_m\}\subset G$ and arbitrary numbers $c_1,\ldots,c_m\in\ccc$,
\begin{equation} \label{pdfncts}
\sum_{j,k}\pd(\gj\gk)\otto \coj\otto\cok \ge 0 \fin , \ \ \ \mbox{($\pd$ is a `positive definite function'~\cite{Folland-AA}).}
\end{equation}

\end{description}
Condition~(\ref{pdfncts}) defining a positive definite function may be regarded as a
`discretization' of~(\ref{clapo-bis}).

Let us now assume that $G$ is \emph{abelian}. By Bochner's theorem~\cite{Folland-AA},
denoting by $\cm(\dug)$ the Banach space of \emph{complex Radon measures} on $\dug$
--- the \emph{dual} group of $G$ --- we can add a further item to the previous list:

\begin{description}

\item[P4\spl] $\pd$ is the Fourier transform of a \emph{positive} measure $\mu$ in $\cm(\dug)$.

\end{description}

The physical relevance of functions of positive type becomes immediately clear once we set
$G=\phasp$ ($\Rightarrow G=\dug$) and we recall that a \emph{classical state} is a normalized positive functional
on the \emph{commutative} $\cast$-algebra of classical observables. This is the algebra
$\cz(\phasp)$ of continuous complex functions \emph{vanishing at infinity},
endowed with the \emph{point-wise product}~\cite{AnielloPF,AnielloFPT}.
The Banach space dual of $\cz(\phasp)$ is $\cm(\phasp)$ and the associated
\emph{states} are the Borel \emph{probability measures} on $\phasp$.
The vector-covector pairing $(f,\mu)\mapsto\langle f\rangle_{\mu}$,
\begin{equation} \label{clev}
\langle f\rangle_{\mu} =  \intrrn f(q,p)\; \de\mu(q,p) \fin , \ \ \
\mbox{with: $f=\overline{f}$ and $\mu$ probability measure,}
\end{equation}
provides the \emph{expectation value} of the \emph{observable} $f$ in the \emph{state} $\mu$.
The latter --- a \emph{set function} --- can conveniently be replaced with an \emph{ordinary function},
its symplectic Fourier transform:
\begin{equation} \label{ftmes}
\pd\qp\equiv\tmu(q,p) \defi \intrrn \exp\tre(\ima\tre\omega(q,p \sei ; \tre q^\prime,p^\prime))
\; \de\mu(q^\prime,p^\prime)\fin , \ \ \
\omega(q,p \sei ; \tre q^\prime,p^\prime) \defi q\cdot p^\prime-p\cdot q^\prime \fin .
\end{equation}
The \emph{characteristic function} $\pd\equiv\tmu$ of $\mu$ is a continuous function of positive type
on $\phasp$. All functions of the latter kind --- that may be called \emph{functions of classical positive type} ---
form a convex cone $\cpdn$. The probability measure normalization condition
$\mu(\phasp)=1$ corresponds to the normalization of $\pd$ as a functional: $\pd(0)=\|\pd\norinf=1$.
Thus, the convex set $\cpdon\subset\cpdn$ of normalized functions of classical positive type
coincides with the set of characteristic functions.

Passing to the quantum setting, recall that in the \emph{phase-space approach} to quantum
mechanics a pure state $\hrho_\psi=|\psi\rangle \langle\psi |$ is replaced with its
\emph{Wigner function} $\rho_\psi\in\lrrn$~\cite{FollandHAPS,Schroeck,AnielloFT,AnielloSP,Wigner},
\begin{equation}
\rho_\psi (q,p) \defi \facn\intrn \mcinque\eee^{-\ima p \cdot x}\otto
\overline{\psi\!\left(q-\frac{x}{2}\right)\!} \dodici
\psi\!\left(q+\frac{x}{2}\right)\dxn \fin ,
\ \ \ \psi\in\ldrn \fin , \ \|\psi\|=1 \fin .
\end{equation}
This (one-to-one) correspondence extends to all trace class operators~\cite{AnielloFT,AnielloSP,AnielloPF,AnielloFPT},
yielding a dense subspace $\wfsn$ of $\lrrn$, containing the convex cone $\pwfsn$
of all functions associated with \emph{positive} trace class operators.
$\pwfsn$ contains the convex set $\pnwfsn$ of all Wigner functions:
\begin{equation} \label{norcon}
\rho\in\pnwfsn \ \Longleftrightarrow \ \rho\in\pwfsn \fin , \
\lim_{r\rightarrow +\infty} \intdisk   \rho(q,p) \; \dqdpn = \tr(\hrho) =1 \fin .
\end{equation}
Here, $\hrho$ is the density operator in $\ldrn$ corresponding to the Wigner function $\rho\in\pnwfsn$.

One can further replace a Wigner function $\rho$ with its \emph{symplectic Fourier-Plancherel transform}
\begin{equation} \label{detsfp}
\big(\fs\sei\rho\big)(q,p)=\frac{1}{(2\pi)^n}\intrrn \otto \rho(q^\prime,p^\prime)\dieci
\ee^{\ima (q\cdot p^\prime - p\cdot q^\prime)}\; \dqdpp \fin ,
\end{equation}
where, in general, the rhs integral should be regarded as a suitable Hilbert space limit~\cite{Aniello-new}.
The subspace $\wfsn$ is mapped by $\fs$  --- a selfadjoint, unitary operator in $\lrrn$ --- onto another
dense subspace of $\lrrn$:
\begin{equation}
\lispn \defi \fs \dodici \wfsn \fin .
\end{equation}
The convex cone $\pwfsn\subset\wfsn$ is mapped by $\fs$ onto a convex cone $\qpdn\subset\lispn$.
By analogy with the classical setting, one may call
\begin{equation} \label{wigetcha}
\tvarrho \defi (2\pi)^n \otto \fs\otto\rho \fin , \ \ \ \mbox{with $\varrho\in\pnwfsn$} \fin ,
\end{equation}
the \emph{quantum characteristic function} of the \emph{quasi-probability distribution}~\cite{AnielloCS} $\rho$.
Similarly to the classical setting, a quantum characteristic function $\tvarrho\in\qpdn$ is characterized by the
normalization condition $\tvarrho(0)=1$. These characteristic functions form a convex subset $\qpdon$ of $\lispn$.
Moreover, one finds out that~\cite{Ali,AnielloFT,AnielloSP}
\begin{equation}
\tvarrho\qp=\tr(U\qp^\ast\hrho)=(\dequm\tre\hrho)\qp \fin ,
\end{equation}
where $U$ is the \emph{Weyl system}; so the previously described group-theoretical dequantization scheme
yields a direct generalization of the characteristic functions rather than of the Wigner functions.

Now, the following problem arises: Is it possible to characterize \emph{intrinsically}
the set $\pnwfsn$ of all Wigner functions or the set $\qpdon$ of quantum characteristic functions?
A rigorous analysis of this problem requires a notion of function of \emph{quantum} positive type,
which, as in the classical setting, relies on a suitable \emph{$\ast$-algebra of functions} and
its \emph{positive functionals}. As previously observed, the Hilbert space $\lrrn$
becomes a $\Hstar$-algebra if endowed with the \emph{twisted convolution}
\begin{equation}
\big(\obse_1\starp \obse_2\big)(q,p)
\defi \frac{1}{(2\pi)^n}\intrrn \obse_1(q^\prime,p^\prime) \,
\obse_2(q-q^\prime,
p-p^\prime)\dieci\eee^{\frac{\ima}{2}(q\cdot p^\prime-p\cdot q^\prime)}
\, \dqdpp \fin ,
\end{equation}
and with a suitable \emph{involution} $\sinvo\colon \obse\mapsto \obse\invo$,
$\obse\invo(q,p) \defi \overline{\obse(-q,-p)}$.
Also recall that the twisted convolution is nothing but the canonical star product associated with
the Weyl system. A \emph{function of quantum positive type} is a positive,
bounded linear functional on the $\Hstar$-algebra $\big(\lrrn, \starp, \sinvo\big)$~\cite{AnielloPF,AnielloFPT}.
Therefore, $\qpt\in\lrrn$ is of quantum positive type if
\begin{equation} \label{quapo}
\intrrn \mcinque\qpt(q,p)\otto (\obse\invo\starp\obse)(q,p)  \; \dqdpn \ge 0 \fin ,  \ \ \
\forall\quattro\obse\in\lrrn \fin . \ \ \ \mbox{(QPTF condition)}
\end{equation}

The functions of quantum positive type --- more precisely, the continuous ones --- enjoy properties
reminiscent of the main properties of their classical counterparts (the functions in $\cpdn$;
i.e., the positive multiples of classical phase-space characteristic functions)~\cite{AnielloPF,AnielloFPT}.
E.g., every \emph{continuous} function of quantum positive type $\qpt$ is bounded and
$\|\qpt\norinf = \qpt(0)$ (recall~{(\ref{nori})}). Moreover, for a \emph{continuous} function
$\qpt\colon\phasp\rightarrow\ccc$ the following facts are equivalent:
\begin{description}

\item[Q1\spl] $\qpt$ is of quantum positive type;

\item[Q2\spl] $\qpt$ is such that ($z\equiv (q,p)\in\phasp$, $\dezn\equiv \dqdpn$,
$\omega(z\tre ,\mtre z^\prime)\equiv q\cdot p^\prime-p\cdot q^\prime$):
\begin{equation} \label{quapo-bis}
\intrdn \intrdn \mcinque\qpt (z-z^\prime)\dieci \overline{\obse(z^\prime)}\otto \obse(z)\dieci
\eee^{\ima\tre \omega(z^\prime\mcinque ,\tre z)/2} \; \dezn \dieci \dezpn \ge 0 \fin ,
\ \ \ \forall\quattro\obse \in \ccom(\phasp) \fin ;
\end{equation}

\item[Q3\spl] for every finite set $\{z_1,\ldots,z_m\}$ in phase space and arbitrary numbers
$c_1,\ldots,c_m\in\ccc$,
\begin{equation} \label{qpdfncts}
\sum_{j,k}\qpt(\zk-\zj)\dieci
\eee^{\ima\tre \omega(\zj\mtre,\tre\zk)/2}\sedici \coj\dodici\cok \ge 0 \fin ;
\ \ \ \mbox{($\qpt$ is a `quantum positive definite function')}
\end{equation}

\item[Q4\spl] $\qpt\in\qpdn$; namely, $\qpt$ is --- up to a normalization factor --- the
(symplectic) Fourier-Plancherel transform of a Wigner distribution.

\end{description}


\section{The physics behind a mathematical \emph{divertissement}}
\label{sect4}


We will now describe an intriguing interplay between functions of \emph{classical} and \emph{quantum}
positive type, and its physical interpretation in the light of the theory of \emph{open quantum systems}.
Once again, the backbone of this construction is group-theoretical.
Recall that the \emph{convolution} product $\muno\convo\mdue$ of two \emph{probability measures}
$\muno,\mdue\in\cm(G)$~\cite{Folland-AA}, is a probability measure too, and indeed the set of
all Radon probability measures on $G$, endowed with this product, becomes a semigroup with identity
(the unit point mass measure $\delta_e$ at the identity $e$ of $G$). If $G$ is \emph{abelian},
the point-wise product $\pd_1\sei\pd_2$ of two (continuous) functions
of positive type on $G$ is a (continuous) function of positive type too, because the Fourier transform
maps the convolution of probability measures into the point-wise multiplication of characteristic functions.
Thus, taking $G=\phasp$, the set $\cpdon\subset\cpdn$ of all normalized functions of `classical' positive type,
endowed with the point-wise product, is a semigroup, with the identity $\pd\equiv 1$.
Now, what is the result of multiplying a function of \emph{classical} positive type
by a continuous function of \emph{quantum} positive type?

\begin{theorem}[\cite{AnielloPF,AnielloFPT}] \label{mainth}
The function $\pd\tre\qpt$, which is obtained by performing the point-wise product of any $\pd\in\cpdn$ by any $\qpt\in\qpdn$,
belongs to $\qpdn$; in particular, if $\pd$ and $\qpt$ are normalized, to the set  $\qpdon$ of quantum characteristic functions.
\end{theorem}

Consider, then, a \emph{multiplication semigroup of functions of classical positive type}
\begin{equation}
\{\clat\colon\phasp\rightarrow\ccc\}_{t\in\errep}\subset\cpdon \fin , \ \ \
\clat\otto\clas=\clats \fin , \ t,s\ge 0 \fin  , \ \ \
\claz \equiv 1 \fin .
\end{equation}
Such semigroups --- assumed to be continuous wrt the the topology of uniform convergence on compact
sets on $\cpdon$~\cite{Folland-AA} --- can be classified, since they are images, via the Fourier transform,
of \emph{convolution semigroups} of probability measures, ruled by the L\'evy-Kintchine formula~\cite{AnielloBM,AnielloQDS}.

The fact that $\clat$ is a bounded continuous function allows us to define the \emph{bounded operator}
\begin{equation} \label{setcsem}
\hcsg_t\colon\lrrn \rightarrow \lrrn \fin , \
\big(\hcsg_t \cinque f\big)(q,p) \defi \clat(q,p)\cinque f(q,p)\fin ,
\ \ \ t\ge 0\fin .
\end{equation}
The set $\{\hcsg_t\}_{t\in\errep}$ is a semigroup of operators~\cite{AnielloBM}.
It is natural to consider a suitable \emph{restriction} of the family of operators $\{\hcsg_t\}_{t\in\errep}$.
Indeed, the complex linear span generated by the convex cone $\qpdn$ of continuous functions of quantum positive type
is the dense subspace $\lispn$ of $\lrrn$, and a semigroup of operators $\{\csg_t\}_{t\in\errep}$ in $\lispn$ is defined as follows.
By Theorem~\ref{mainth}, it is consistent to set~\cite{AnielloPF,AnielloFPT}
\begin{equation} \label{setcsem-bis}
\csg_t\colon\lispn\rightarrow\lispn\fin , \ \ \
\big(\csg_t \qpt\big)(q,p) \defi \clat(q,p)\tre \qpt(q,p)\fin ,
\end{equation}
and we have: $\csg_t \cinque \qpdn\subset\qpdn$, $\csg_t \cinque \qpdon\subset\qpdon$.
$\{\csg_t\}_{t\in\errep}$ is called a \emph{classical-quantum semigroup}~\cite{AnielloCQS}.

A classical-quantum semigroup is not only a mathematical \emph{divertissement}. The Weyl system $U$
induces a symmetry action of the group $\phasp$ on the space $\trc$ (with $\hh=\ldrn$), where
the quantum states live; i.e., the \emph{isometric representation}  $\urep$, where
$\urep\qp\sei A\defi U\qp\sei A \otto U\qp^\ast$. Given a convolution semigroup
$\{\mut\}_{t\in\errep}$ of probability measures on $\phasp$, setting
\begin{equation} \label{deunim}
\unimu\sei A \defi \intrrn  \big(\urep\qp\sei A\big)\dieci \de \mut\qp \fin ,
\end{equation}
one obtains a semigroup of operators $\{\unimu\}_{t\in\errep}$ in $\trc$,
a \emph{twirling semigroup}~\cite{AnielloBM,AnielloQDS,AnielloCQS,AnielloSO,Aniello-OvF,AnielloCD}.

\begin{theorem}[\cite{AnielloPF,AnielloFPT,AnielloBM,AnielloCQS,AnielloSO,AnielloCD,Aniello-SchCo}]
Let $\{\clat\}_{t\in\errep}$ be the multiplication semigroup of functions of classical positive type such that
$\clat(q,p)=\int \eee^{\ima (q\cdot p^\prime - p\cdot q^\prime)}\; \de\mut(q^\prime,p^\prime)$,
and let $\{\csg_t\}_{t\in\errep}$ be the associated classical-quantum semigroup.
Then, the quantization map $\qum\colon\lrrn\rightarrow\hs$, generated by the Weyl system $U$,
intertwines $\{\csg_t\}_{t\in\errep}$ with the twirling semigroup $\{\unimu\}_{t\in\errep}$, namely,
\begin{equation} \label{interqum}
\qum\sei(\csg_t \qpt) = \unimu\sei(\qum\tre\qpt) \fin , \ \ \ \qpt\in\lispn
\ (\Leftrightarrow \sei \qum\tre\qpt\in\trc)\fin , \ \ t\ge 0 \fin .
\end{equation}
$\{\unimu\}_{t\in\errep}$ is a quantum dynamical semigroup --- i.e., a semigroup of completely positive,
trace-preserving linear maps --- thus it describes the temporal evolution of an open quantum system.
Moreover, it does not decrease the von~Neumann entropy of a quantum state.
\end{theorem}


\section{State-preserving products from square integrable representations}
\label{sect5}


Quantum \emph{observables} are embedded in a $\mathrm{C}^\ast$-algebra, the bounded operators, with the algebraic structure
provided by the product (composition) of operators. \emph{States} --- density operators: unit trace, positive trace class operators
--- are only `indirectly' involved in this structure, as positive functionals on this algebra. The product of two states is not,
in general, itself a state. Endowing the space of trace class operators with, e.g., the \emph{Jordan product}
$(A,B)\mapsto A\jop B\defi(AB+BA)/2$, or, say, with the \emph{Lie product} $(A,B)\mapsto A \lip B \defi(AB-BA)/2\ima$,
one gets algebraic structures preserving selfadjointness. But the Jordan product is not associative,
and the composition or the Jordan product of two states is itself a state if and only if the two factors are equal
and \emph{pure} (i.e., rank-one projectors); whereas the Lie product of two states is \emph{never} a quantum state.
Precisely, let $\hrho$, $\hsigma$ be \emph{density operators} in a separable complex Hilbert space $\hh$;
then:
\begin{enumerate}[(i)]

\item \label{comp}
$\hrho\otto\hsigma$ is a density operator if and only if $\hrho=\hsigma\equiv \hpi$, where $\hpi$ is a pure state;

\item  \label{jord}
$\hrho\jop\hsigma$ is a density operator if and only if $\hrho=\hsigma\equiv \hpi$, where $\hpi$ is a pure state;

\item  \label{liep}
$\hrho\lip\hsigma$ is not a density operator.

\end{enumerate}
The ``if part'' of~(\ref{comp}), (\ref{jord}) is obvious.
Moreover, denoting by $\langle\cdot\sei,\cdot\ranglehs$ the Hilbert-Schmidt product,
\begin{equation}
\tr(\hrho\otto\hsigma)=\langle \hrho,\hsigma\ranglehs\le\sqrt{\tr\bile \hrho^2\biri}\otto\sqrt{\tr\bile \hsigma^2\biri} \fin ,
\ \ \ \mbox{(Cauchy-Schwarz inequality)}
\end{equation}
and $\tr\bile \hrho^2\biri<1$ if $\hrho$ is not pure. Hence, $\tr(\hrho\otto\hsigma)<1$ if either $\hrho$ or $\hsigma$ is not pure;
or if they are both pure, but do not coincide. This proves the first assertion. To prove the second assertion, just notice that
$\tr(\hrho\jop\hsigma)=\tr(\hrho\otto\hsigma)$. Hence, by the previous point, $\tr(\hrho\jop\hsigma)=1$ if and only if
$\hrho$, $\hsigma$ are pure states and $\hrho=\hsigma$. Finally, $\hrho\lip\hsigma$ cannot be a density operator because
$\tr(\hrho\otto\hsigma-\hsigma\sei\hrho)=0$.

Now, one can ask: Is it possible to endow the Banach space $\trc$ with a binary operation
giving rise to an associative algebra structure and such that the product of two states is a state too?
The answer is positive, and to construct such a product we need the following ingredients:
\begin{itemize}

\item A \emph{square integrable} projective representation $U$ of a locally compact group $G$ in $\hh$.
For simplicity, we assume that $G$ is \emph{unimodular} (recall the positive constant $\ddu$ in Theorem~\ref{Duflo-Moore}).

\item A \emph{fiducial density operator} $\htau\in\trc$. The choice of ($U$ and) $\htau$ characterizes the product.

\end{itemize}
Then, we set ($\hame$ Haar measure, $\urep(g)\sei \htau\defi U(g)\sei \htau \otto U(g)^\ast$ and Bochner integral on the rhs):
\begin{equation} \label{twipro}
A \pro B \defi d_U^{-2} \intG \de\hame(g) \nove \tr\bile A\sei (\urep(g) \sei \htau)\biri\sei (\urep(g) \tre B)\in\trc \fin ,
\ \ \ \mbox{(`twirled product')}
\end{equation}
for all $A,B\in\trc$. One can prove that $(\trc,\pro)$ is an \emph{associative algebra}, and the twirled product
$\pro$ is \emph{state-preserving} (if $\hrho$ and $\hsigma$ are density operators, $\hrho \pro \hsigma$ is a density operator too).
The square-integrability of the representation $U$ is an essential ingredient here, because with this condition
$d_U^{-2}\tr(A\sei (\urep(g) \sei \htau))\sei\de\hame(g)$ is a complex measure (actually, a probability measure if $A\equiv\hrho$ is a
density operator) --- see Proposition~{7} of~\cite{AnielloFT} --- so that definition~{(\ref{twipro})} is consistent.



\end{document}